\documentclass[twocolumn,10pt]{article}

\newbox\tempbox
\newenvironment{nomenclature}{%
   \newcommand\entry[2]{%
       \setbox\tempbox\hbox{##1.\quad}
       \hangindent\wd\tempbox\noindent{##1}\quad\ignorespaces##2\par}
       \section*{Nomenclature}}{\par\addvspace{12pt}}

\AtEndOfClass{%
  \def\@begintheorem#1#2{\trivlist \item[\hskip \labelsep%
  {\tensfb #1\ #2.}]\it}
  \def\@opargbegintheorem#1#2#3{\trivlist
        \item[\hskip \labelsep%
  {\tensfb #1\ #2\ (#3).}]\it}
  \def\newtheorem{\@ifstar{\@newstheorem}{\@newtheorem}}
  \def\@newtheorem#1{\@ifnextchar[{\@othm{#1}}{\@nthm{#1}}}
  \def\@newstheorem#1#2{\@namedef{#1*}
  {\@ifnextchar[{\@opargbeginstheorem{#2}}{\@beginstheorem{#2}}}%
  \@namedef{end#1*}{\@endtheorem}}
  \def\@beginstheorem#1{\trivlist \item[\hskip \labelsep%
  {\tensfb #1.}]\it}
  \def\@opargbeginstheorem#1[#2]{\trivlist\item[\hskip \labelsep%
  {\tensfb #1\ (#2).  }]\it}
  \def\proof{\@ifnextchar[{\@optargproof}{\@proof}}
  \def\@proof{\trivlist \item[\hskip \labelsep{\it Proof.}]}
  \def\@optargproof[#1]{\trivlist \item[\hskip \labelsep{\it #1.}]}
  
  \def\qed{\vskip-\lastskip\vskip-\baselineskip\hbox to \hsize{\hfill$\Box$}} 
}

\usepackage[margin=0.75in]{geometry}
\usepackage{times,newtxmath}
\usepackage[T1]{fontenc}
\usepackage[utf8]{inputenc}
\usepackage[small]{titlesec}

\usepackage{color}
\usepackage{graphicx} 
\usepackage{amssymb,amsmath,bm} 

\usepackage{fancyhdr}
\usepackage{lastpage}
\usepackage{hyperref} 

\title{On the enhancement of heat transfer and reduction of entropy generation by asymmetric slip in pressure-driven non-Newtonian microflows}

\author{\textbf{\textsf{Vishal Anand}}\\
        Graduate Research Assistant\\
       \textbf{\textsf{Ivan C.\ Christov}}\footnote{Author to whom correspondence should be addressed.}\\
        Assistant Professor\\[1mm] 
        School of Mechanical Engineering\\
        Purdue University\\
        West Lafayette, Indiana, 47907\\
        Email: christov@purdue.edu
}

\date{}

\begin{document}
\maketitle    

\begin{abstract}
{\it We study hydrodynamics, heat transfer and entropy generation in pressure-driven microchannel flow of a power-law fluid. Specifically, we address the  effect of asymmetry in the  slip boundary condition at the channel walls. Constant, uniform but unequal heat fluxes are imposed at the walls in this thermally developed flow. The effect of asymmetric slip on the velocity profile, on the wall shear stress, on the temperature distribution, on the Bejan number profiles, and on the average entropy generation and the Nusselt number are established through the numerical evaluation of exact analytical expressions derived. Specifically, due to asymmetric slip, the fluid momentum flux and thermal energy flux are enhanced along the wall with larger slip, which in turn shifts the location of the velocity's maximum to an off-center location closer to the said wall. Asymmetric slip is also shown to redistribute the peaks and plateaus of the Bejan number profile across the microchannel, showing a sharp transition between entropy generation due to heat transfer and due to fluid flow at an off-center-line location. In the presence of asymmetric slip, the difference in the imposed heat fluxes leads to starkly different Bejan number profiles depending on which wall is hotter, and whether the fluid is shear-thinning or shear-thickening. Overall, slip is shown to promote uniformity in both the velocity field and the temperature field, thereby reducing irreversibility in this flow.}
\end{abstract}



\section{Introduction}
\label{S:1}

Contrary to what we teach in undergraduate fluid mechanics, the velocity of a fluid at a surface may or may not be equal to the velocity of said surface. This discontinuity in velocity at a fluid-solid interface is known as \emph{slip}. Slip flow is encountered in industrial hydrodynamic processes across many length scales \cite{Lauga2007}. At the macro-scale, slip flow occurs in extrusion of polymers, wherein it is caused by stress instabilities, and affects the quality of final product \cite{Denn1990}. At the micro- and nanoscale, slip flow is even more common \cite{GEH99}. For microscale flow of rarefied gases, a dimensionless number, termed the \emph{Knudsen number}, which is the ratio of the mean free path of the gas molecules to a characteristic channel dimension, determines the flow regime. When $0.001 <Kn <0.1$, the flow can still be treated in a continuum sense as long as slip at boundaries (``slip flow'') is allowed \cite{NW06}. Here, the slip velocity boundary condition may also be accompanied by a temperature jump boundary condition, wherein there is a discontinuity in temperature at the fluid-solid interface \cite{GEH99,STC17}. Although less common than in gases, slip flow may still be observed in microflows of liquids, when the surface of the channel is made of a hydrophobic material, or when the surface has microscale features such as roughness \cite{Lauga2007}.

At the macro-scale, on the other hand, Denn \cite{Denn1990,Denn2001ExtrusionSlip} provides a review of slip flows of non-Newtonian fluids, specifically polymer melts, while Kalyon \cite{Kalyon} discusses the apparent slip of concentrated suspensions. On the basis of molecular dynamics simulations, Thompson and Troian \cite{Thompson1997} substantiated a generalized \emph{nonlinear} slip boundary condition in which the slip length is a function of the shear rate, thereby generalizing the original slip law of Navier \cite{Navier}. More recently, Cloitre and Bonnecaze \cite{Cloitre2017} summarized the experimental approaches and evidence for nonlinear slip in flows of soft materials. Mathews and Hill \cite{Matthews2007} initiated the theoretical work on Newtonian fluids subject to  a nonlinear slip boundary condition by finding, analytically, the velocity profile for three canonical pressure-driven flows: in a pipe, in an annulus and in a channel. 

From the early works of Bird, Acrivos et al.~\cite{Bird1959,Acrivos1960} on flow and heat transfer of non-Newtonian fluids to the more ``exotic'' modern applications such as flow in self-affine subsurface fractures \cite{YanKoplik2008}, the power-law fluid model (also known as the Ostwald--de Waele model) is a good starting point for modeling non-Newtonian fluid behavior, in part due to the power-law fluid's well-characterized material properties. Therefore, we restrict ourselves to this class of non-Newtonian fluids, which Bird featured as one of several selected ``useful non-Newtonian models'' \cite{Bird76}. Thus, relevant to our present work is the analytical and semi-analytical study by Ferras et al.~\cite{Ferras2012AnalyticalSlip} of the Couette--Poiseuille flow of Newtonian and generalized Newtonian fluids. Specifically, they obtained mathematical expressions for the velocity profiles for different types of inelastic fluids (e.g., power-law and Bingham fluids), for pressure-driven flows with either imposed pressure drop or imposed flow rate. Meanwhile, Pritchard et al.~\cite{Pritchard2011} provide an overview and some new results regarding the development of \emph{unsteady} unidirectional flow of power-law fluids, specifically Stokes' second problem of the oscillating plate. More recently, Wei and Jordan \cite{WeiJordan2013} derived exact solutions for various traveling wave forms in \emph{compressible} power-law fluids.  

Transport in microchannels is an active area of research, especially in the heat transfer community, in particular due to the numerous applications related to the cooling of microelectronics \cite{gs03}. However, much of the research summarized in the authoritative reviews \cite{gs03,Yovanovich2015} is on Newtonian working fluids, though slip effects are mentioned to be important in some flow regimes. On the other hand, though non-Newtonian fluids play a critical role in unconventional energy applications such as hydraulic fracturing \cite{AnnRevFrack}, wherein slip and shear-thinning/thickening can be observed, thermal effects have not been addressed in detail. As for the heat transfer characteristics of non-Newtonian fluids, Barletta \cite{BarlettaFullyDissipation} proposed a new expression for the Brinkman number for power-law fluids, which is a dimensionless number that quantifies viscous dissipation, in his study of fully developed laminar convection of a power-law fluid with viscous dissipation. The Nusselt number correlations (and their method of computation) in laminar heat transfer of non-Newtonian fluids is of practical interest~\cite{Cruz2012}. Jambal et al.~\cite{Jambal2005} incorporated the effect of axial wall conduction in laminar convection of a power-law fluid with uniform wall temperature, deriving scalings for the Nusselt and Brinkman numbers in the thermal entrance region. Tso et al.~\cite{Tso2010} considered the effect of asymmetric heat flux and viscous dissipation on heat transfer characteristics of fully developed flow of power-law fluids, showing that, for the case of unequal heat fluxes on the microchannel walls, the Nusselt number depends the Brinkman number. Meanwhile, Sheela-Fransisca et al.~\cite{Sheela-Francisca2012} derived mathematical expressions relating the Nusselt number to the Brinkman number in the asymmetric thermal viscous-dissipative Couette--Poiseuille flow of a pseudo-plastic fluid. Additionally, Straughan \cite[Chap.~9]{StraughanNonEqBook} provides an excellent overview of the effects of the Navier slip condition on various heat transfer phenomena in Newtonian fluids. Kaushik et al.~\cite{Kaushik2016} studied numerically the fluid dynamics, heat transfer and entropy generation characteristics of extrusion flows of power-law fluids between parallel plates under the lubrication approximation in quasi-steady state. They concluded that Nusselt number increases/decreases with time for shear-thinning/thickening fluids. For shear-thickening fluids, they observed unusual negative values of the Nusselt number. Sefid et al.~\cite{Sefid2013} presented numerical results on the thermal characteristics of developing and developed flow of power-law fluids in concentric annuli, demonstrating the entrance length's increases with the power-law index.

Beyond the heat transfer, it is of interest to also quantify the amount of entropy generated in a fluid system because, according to the Gouy--Studola theorem, the latter determines the work lost in the thermomechanical process \cite{BejanAdr1999}. At the microscale, the velocity and temperature gradients encountered are larger than at the macroscale, which motivates us to additionally undertake the analysis of entropy generation of such thermal systems \cite{BejanAdr1999}. Previously, Mahmud and Fraser \cite{Mahmud2002,Mahmud2006} minimized the entropy generated as a function of the power-law index in the flow between parallel plates \cite{Mahmud2002} and in a circular tube \cite{Mahmud2006}. Hung \cite{Hung2008} extended the latter to include viscous dissipation, showing that the rate of entropy generation increases with the Brinkman number, and this enhancement is more pronounced for shear-thickening fluids. Shojaeian et al.~\cite{Shojaeian2014ConvectiveConditions} studied the effect of slip flow (under the \emph{linear} Navier slip wall) on heat transfer and entropy generation characteristics of Newtonian and power-law fluids in microchannels. They found that the Nusselt number increases with slip, and this increase is, once again, more prominent for shear-thickening than shear-thinning fluids. Anand \cite{Anand2014} studied how the choice of a slip law influences the entropy generation and heat transfer characteristics of pressure-driven flow of a power-law fluid. He considered three slip laws for his study, namely, Navier's slip law, Hatzikiriakos' slip law, and the asymptotic slip law. He showed that, for the same slip coefficient, the Hatzikiriakos slip law predicts a higher Nusselt number and less entropy generation compared to the asymptotic slip law. Goswami et al.~\cite{Goswami2016} studied the entropy generation minimization of an electrosmotic flow of a power-law fluid taking into account the conjugate heat transfer. They derived approximate analytical solutions and concluded that the entropy generation has an optimum (minimum) value for a certain combination of wall thickness, wall thermal conductivity, and Biot number. Meanwhile, Mondal \cite{M14} studied entropy generation in the combined Poiseuille--Couette flow of a power-law fluid between two asymmetrically heated plates. He found that  irreversibility in shear-thinning fluids is primarily due to fluid friction, while it is primarily due to heat transfer in shear-thickening fluids. Viscous heating was also considered, finding that the Nusselt number suffers a singularity for negative Brinkman numbers due to a balance between the viscous dissipation in the fluid by heat transfer from the walls.

The existing literature dealing with slip flows assumes that the slip is identical on both walls of a channel. However, such symmetry is not necessarily always present. If the two channel walls are made from different materials (e.g., one is made from a polymer such as polydimethylsiloxane (PDMS), while the other is made from glass) or each has a different surface roughness, then asymmetric slip occurs. A similar situation occurs in the subsurface, wherein non-Newtonian fluids are pumped through complex fractured rocks with highly heterogeneous surface properties \cite{AnnRevFrack}. Vayssade et al.~\cite{Vayssade2014} performed an experimental study of suspension flow in a microchannel, providing evidence for asymmetric slip velocities along the channel walls. Panaseti et al.~\cite{Georgi1} extended the latter approach, which was based on the Herschel--Bulkley (HB) fluid model, to account for slip via Navier's nonlinear slip condition. However, thermal effects in such flows have not been considered.

The present work aims to address the latter apparent gap in the archival literature. Here, we address the effect of asymmetric slip boundary conditions on the hydrodynamics, heat transfer and the entropy generation characteristics of thermally fully developed flow of a power-law fluid in a microchannel under uniform heat flux thermal boundary conditions. Following  \cite{Ferras2012AnalyticalSlip,Anand2014}, we implement the slip boundary condition using Navier's nonlinear slip law \cite{Matthews2007}. We show, through a comparative analysis in Sect.~\ref{sec:results}, that asymmetry in the slip velocities at the channel walls has a significant influence on the fluid flow and the heat transfer characteristics of the problem, specifically on the enhancement of heat transfer and the reduction of irreversibility as compared to the usual case of symmetric slip boundary conditions. Our approach is analytical, based on the exact results that we first obtain in Sect.~\ref{S:2} for the fluid flow, heat transfer and entropy generation in a microchannel with asymmetric imposed heat fluxes and unequal wall slip coefficients.

\section{Problem statement and mathematical analysis}
\label{S:2}

\begin{figure}[ht]
\centering\includegraphics[width=0.9\linewidth]{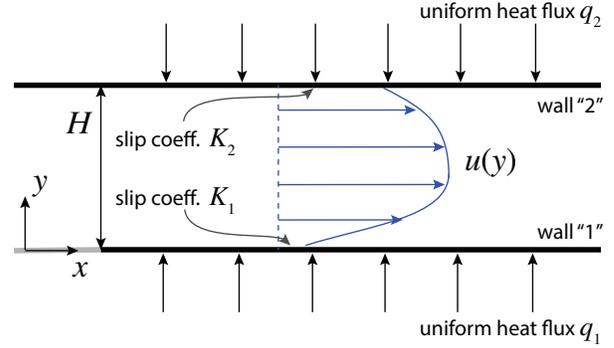}
\caption{Schematic of the physical model, coordinates and notation. We work the problem per unit width out of the page (i.e., the $+z$-direction).} 
\label{scheme}
\end{figure}

As Stone notes, ``the equations [of fluid mechanics] are generally difficult to solve, (rational) approximations are necessary and frequently the interplay of physical arguments, mathematical simplifications, and experimental insights are crucial to progress and understanding'' \cite{Stone2017}. To this end, we would like to pose a tractable problem of non-Newtonian fluid flow and heat transfer in the presence of asymmetric wall slip. Specifically, we consider the pressure-driven flow in a microchannel as shown in Fig.~\ref{scheme}. Although a ``model flow'' from the engineering point-of-view, understanding this basic flow is of fundamental interest as it is relevant to any context in which there is a separation of scales between a ``small'' cross-sectional and a ``long'' flow-wise direction (see, e.g., the discussion in \cite[Sect.~1.8.2]{Stone2017}). 

In Fig.~\ref{scheme}, the microchannel is of height $H$, the lower wall is labeled ``1,'' and the upper wall is labeled ``2.'' The flow is maintained by constant pressure gradient $G$.\footnote{To be precise, $G$ is the inlet--outlet pressure difference per unit length of channel.} The lower and upper walls are each subject to a uniform and constant heat flux $q_1$ and $q_2$, respectively. The slip coefficient, the meaning of which will be made clear through Eq.~\eqref{velocity_BC} below, on the lower wall is $K_1$, while  on the upper wall it is $K_2$. The $x$-axis is along the lower wall of the channel, in the flow-wise direction, while the $y$-axis is perpendicular to the lower wall, in the span-wise direction.

\subsection{Stress distribution}

For a steady unidirectional flow in the $x$-direction, specifically $\bm{v}=\big(u(y),0,0\big)$, the conservation of linear momentum becomes \cite{Leal}:
\begin{equation}
\label{stress}
\frac{d\tau_{yx}}{dy} = -G,
\end{equation}
where $G$ is the pressure gradient and $\tau_{yx}$ is the relevant component of the shear stress tensor $\bm{\tau}$. Alternatively, restricting to flow in a microchannel, the inertial forces can be neglected in comparison to the viscous and pressure forces, which once again yields Eq.~\eqref{stress}. 

Integrating Eq.~\eqref{stress} with respect to $y$, we obtain
\begin{equation}
\label{stress2}
\tau_{yx} = -Gy + C, 
\end{equation}
where $C$ is an integration constant to be determined. At the lower wall (wall ``1'' at $y=0$), the shear stress is given by $\tau_{w1}$, i.e., $\tau_{yx}|_{y=0} =\tau_{w1}$. Then, $C=\tau_{w1}$ and Eq.~\eqref{stress2} becomes
\begin{align}
\label{stress3}
\tau_{yx} = -Gy + \tau_{w1},
\end{align}
whence the shear stress on the top wall (wall ``2'' at $y = H$) is  found to be
\begin{equation}
\label{stress_tw}
\tau_{w2} \equiv \tau_{yx}|_{y=H} = -GH + \tau_{w1}.
\end{equation}
In what follows, it will be necessary to know the vertical location, $y = y_0$, in the span-wise direction of the channel at which $\tau_{yx}=0$ (i.e., where the shear stress vanishes and the velocity gradient changes sign). This location is obtained by setting the right-hand side of Eq.~\eqref{stress3} to zero, which yields
\begin{equation}
\label{y_0}
y_0=\tau_{w1}/G.
\end{equation}

The slip boundary condition can be understood as ``coupling between the local velocity and the surface stresses'' \cite[Sect.~1.4.1]{Stone2017}. Specifically, the slip velocity at a given wall can be expressed in terms of the  shear stress at that wall. Following \cite{Matthews2007,Ferras2012AnalyticalSlip}, we employ Navier's nonlinear slip law. The slip velocity at wall ``1'' and wall ``2'' are then denoted by $u_{w1}$ and $u_{w2}$, respectively. Thus, the velocity--stress relation at the channel walls takes the form (see also \cite[Eq.~(5)]{Cloitre2017} and the discussion thereof):
\begin{subequations}\begin{align}
u_{w1}&=(\tau_{w1})^mK_1, \\
\label{velocity_BC2}
u_{w2}&=(-\tau_{w2})^mK_2 
=(GH-\tau_{w1})^mK_2,
\end{align}\label{velocity_BC}\end{subequations}
where $K_1$ and $K_2$ are the \emph{slip coefficients} and $m$ is the \emph{slip exponent}. For $m=1$, Navier's (linear) slip law \cite{Navier,Lauga2007} is obtained. The slip coefficients and the slip exponent depend on various physical factors such as dissolved phases in the fluid, wall surface roughness and wetting properties, electro-rheological behavior of the fluid, etc.~\cite{Lauga2007}. In the present work, we consider the slip to be asymmetric, i.e., $K_1 \ne K_2$, but we assume that the slip exponent $m$ is the same on both walls (see, e.g., \cite{Vayssade2014}).

\subsection{Velocity distribution}
\label{sec:vel}

Next, we seek to determine the velocity distribution across the microchannel, which is sketched representatively in Fig.~\ref{scheme}. To that end, we first need to prescribe the {constitutive equation} for the fluid. In the present work, following  \cite{Anand2014,Ferras2012AnalyticalSlip,Chhabra2010}, we consider the fluid to be non-Newtonian and its shear stress--strain rate relationship to be well-described by the power-law model \cite{Bird}.  For unidirectional flow, the latter relationship takes the form
\begin{equation}
\label{power_law}
\tau_{yx} = a\left|\frac{du}{dy}\right|^{n-1}\frac{du}{dy}.
\end{equation}
Here, $a$ is the \emph{consistency factor}, while $n$ is the \emph{power-law index}. According to the power-law model, when $n <1$, the apparent viscosity, $\eta = a |du/dy|^{n-1}$, of the fluid decreases with the shear rate-of-strain, and such a fluid is termed \emph{shear-thinning}. On the other hand,  when $n>1$, the apparent viscosity of the fluid increases with the shear rate-of-strain, and such a fluid is termed \emph{shear-thickening}. The power-law model in Eq.~\eqref{power_law} reduces to Newton's law of viscosity for $n=1$, in which case $a$ becomes the usual shear viscosity.

In the lower portion of the microchannel, $y < y_0$ and $du/dy >0$, which allows us to write Eq.~\eqref{power_law} for shear stress as
\begin{equation}
\tau_{yx}=a\left(\frac{du}{dy}\right)^n.
\end{equation}
Substituting the latter into Eq.~\eqref{stress3},  yields the following first-order ordinary differential equation (ODE) for $u(y)$:
\begin{equation}
a\left(\frac{du}{dy}\right)^n = -Gy +\tau_{w1},
\end{equation}
The latter ODE is easily integrated to give
\begin{equation}
u(y) = C_1 - \frac{(\tau_{w1}-Gy)^{1/n+1}}{(1/n+1)a^{1/n}G},
\end{equation}
where $C_1$ is an arbitrary constant of integration. The boundary condition $u(0)=u_{w1}$ allows us to solve for the integration constant:
\begin{equation}
\label{C_1}
C_1 = u_{w1}+\frac{(\tau_{w1})^{1/n+1}}{(1/n+1)a^{1/n}G}.
\end{equation}
Thus, the expression for velocity distribution in the strip $0 \le y < y_0$ is
\begin{equation}
\label{vd_lower}
u(y) = u_{w1} + \frac{(\tau_{w1})^{1/n+1}}{(1/n+1)a^{1/n}G} \left[1 - \left(1-\frac{Gy}{\tau_{w1}}\right)^{1/n+1} \right].
\end{equation}

In the upper portion of the microchannel, $y> y_0$ and $du/dy < 0$, which allows us to, now, write Eq.~\eqref{power_law} for  the shear stress as
\begin{equation}
\tau_{yx} = - a\left(-\frac{du}{dy}\right)^n.
\end{equation}
Substituting the latter into Eq.~\eqref{stress3}, again yields a first-order ODE for $u(y)$:
\begin{equation}
a\left(-\frac{du}{dy}\right)^n=Gy-\tau_{w1},
\end{equation}
which is easily integrated to obtain
\begin{equation}
u(y) = -\frac{(Gy-\tau_{w1})^{1/n+1}}{(1/n+1)a^{1/n}G} + C_2.
\end{equation}
The integration constant $C_2$ is set by the  boundary condition at wall 2, i.e., $u(H) = u_{w2}$:
\begin{equation}
C_2 = u_{w2} + \frac{(GH-\tau_{w1})^{1/n+1}}{(1/n+1)a^{1/n}G}.
\end{equation}
Thus, the expression for velocity distribution in the strip $y_0 < y \le H$ is
\begin{multline}
\label{vd_upper}
u(y) = u_{w2} + \frac{(\tau_{w1})^{1/n+1}}{(1/n+1)a^{1/n}G} \Bigg[ \left(\frac{GH}{\tau_{w1}}-1\right)^{1/n+1} \\ - \left(\frac{Gy}{\tau_{w1}}-1\right)^{1/n+1} \Bigg].
\end{multline}

All quantities of interest have been written in terms of $\tau_{w1}$, which is the only unknown remaining now. To find an expression for $\tau_{w1}$, we must require that the velocity profile $u(y)$ be continuous across $y=y_0$:
\begin{equation}
\label{continuity}
\lim_{y\to y_0^-} u(y) = \lim_{y\to y_0^+} u(y).
\end{equation}
Thus, upon substituting Eq.~\eqref{y_0} for $y_0$ and Eqs.~\eqref{velocity_BC} for $u_{w1}$ and $u_{w2}$ into Eqs.~\eqref{vd_lower} and \eqref{vd_upper}, we obtain
\begin{multline}
\label{continuity3}
(\tau_{w1})^mK_1 + \frac{(\tau_{w1})^{1/n+1}}{(1/n+1)a^{1/n}G}
= (\tau_{w1})^m\left(\frac{GH}{\tau_{w1}}-1\right)^m K_2 \\
+\frac{(\tau_{w1})^{1/n+1}}{(1/n+1)a^{1/n}G} \left(\frac{GH}{\tau_{w1}}-1\right)^{1/n+1}.
\end{multline}
Equation~\eqref{continuity3} is an \emph{implicit, nonlinear} algebraic equation for $\tau_{w1}$ given $K_1$, $K_2$, $G$, $a$, $n$ and $H$. In practice, it must be inverted numerically. Nevertheless, this completes the derivation of the unidirectional velocity profile $u(y)$ of a power-law fluid in a microchannel with asymmetric slip.  To summarize: $u(y)$ is piecewise defined by Eqs.~\eqref{vd_lower} and \eqref{vd_upper} and $\tau_{w1}$ is found by solving Eq.~\eqref{continuity3} numerically. 

From the above equations, some characteristic physical scales are evident. The characteristic length scale in the problem is $\ell_c=H$. The pressure gradient then sets a character stress scale $\mathcal{T}_c = G\ell_\mathrm{c} = GH$ and a characteristic  velocity scale is evident from Eqs.~\eqref{vd_lower} and \eqref{vd_upper}: $\mathcal{V}_c = \mathcal{T}_c^{1/n+1}/[(1/n+1)a^{1/n}G]$. 
Hence, we are led to introduce the following dimensionless variables:
\begin{alignat}{3}
\label{length_scale}
\bar{y} &= y/\ell_c && = y/H, \\
\label{stress_scale}
\bar{\tau}_{w}(\bar{y}) &= \tau(y)/\mathcal{T}_c && = \tau(y)/(GH), \\
\label{velocity_scale}
\bar{u}(\bar{y}) &= u(y)/\mathcal{V}_c && = u(y)\, \frac{(1/n+1)a^{1/n}G}{(GH)^{1/n+1}}, \\
\label{K_scale}
\bar{K}_{1,2} &= K_{1,2} (\ell_c^m / \mathcal{V}_c) && = K_{1,2} \, \frac{(1/n+1)a^{1/n}G}{(GH)^{1/n+1-m}}.
\end{alignat}
Using the latter set of dimensionless variables, we obtain a set of dimensionless, coupled nonlinear algebraic equations that fully specify the velocity profile:
\begin{equation}
\label{velocity_nd}
\bar{u}(\bar{y}) = \begin{cases} \bar{u}_{w1}+(\bar{\tau}_{w1})^{1/n+1}-(\bar{\tau}_{w1}-\bar{y})^{1/n+1}, & \bar{y} \le \bar{\tau}_{w1},\\
\bar{u}_{w2}-(\bar{y}-\bar{\tau}_{w1})^{1/n+1}+(1-\bar{\tau}_{w1})^{1/n+1}, & \bar{y} > \bar{\tau}_{w1}, 
\end{cases}
\end{equation}
where Navier's nonlinear slip law gives
\begin{subequations}\begin{align}
\label{velocity_nd_bc1}
\bar{u}_{w1} &=\bar{K}_1(\bar{\tau}_{w1})^m, \\
\label{velocity_nd_bc2}
\bar{u}_{w2} &=\bar{K}_2(1-\bar{\tau}_{w1})^m,
\end{align}\label{velocity_nd_bc}\end{subequations}
and $\bar{\tau}_{w1}$ satisfies
\begin{equation}
\label{velocity_nd_tau_eq}
\bar{K}_1(\bar{\tau}_{w1})^m+(\bar{\tau}_{w1})^{1/n+1} = \bar{K}_2(1-\bar{\tau}_{w1})^m+(1-\bar{\tau}_{w1})^{1/n+1}.
\end{equation}
As a consistency check, note that, for the special case of symmetric slip with $\bar{K_1}=\bar{K_2}$, $\bar{\tau}_{w1}=1/2$ is obviously a solution of Eq.~\eqref{velocity_nd_tau_eq}. Therefore, for symmetric slip, the velocity gradient vanishes exactly along the channel's centerline.

Note that the velocity profile for a power-law fluid with asymmetric slip developed above, i.e., Eqs.~\eqref{velocity_nd}--\eqref{velocity_nd_tau_eq}, was also considered in \cite[Eq.~(23)--(25)]{Georgi1}, wherein the power-law fluid is a special case of a derivation for HB fluids. Of course, in showing an equivalence, one has to take into account the different characteristic scales employed in \cite{Georgi1}.

\subsection{Temperature distribution}
\label{sec:temp}

The energy equation for the thermally developed flow of a power-law fluid in a microchannel (see, e.g., \cite{Hung2008,Anand2014} and the references therein) is given by
\begin{equation}
\label{energy}
\rho c_p u(y)\frac{\partial T}{\partial x}= k\frac{\partial ^2T}{\partial y^2}+ a\left|\frac{du}{dy}\right|^{n-1}\left(\frac{du}{dy}\right)^2,
\end{equation}
where viscous dissipation cannot be neglected for microflows, as discussed by Koo and Kleinstreuer \cite{KK04}. 
Here  $\rho$, $c_p$, $k$ are respectively the density, specific heat and thermal conductivity of the fluid, which are assumed constant for this problem, and $T(x,y)$ is the temperature profile in the microchannel. 
Let us introduce the dimensionless temperature
\begin{equation}
\label{theta}
\theta(\bar{y}) = \frac{k}{q_1H}\big[T(x,y)-T_{w1}(x)\big],
\end{equation}
where, following our convention for the velocity profile, we denote the temperature profiles on the lower and upper channel walls as $T_{w1}(x) \equiv T(x,0)$ and $T_{w2}(x) \equiv T(x,H)$, respectively. We also note that although $\partial T/\partial x\ne0$ [see Eqs.~\eqref{gamma0} and \eqref{gamma} below],  ${\partial\theta}/{\partial x}=0$, i.e., $\theta=\theta(\bar{y})$ only, because of the assumption of thermally developed flow under  uniform heat flux boundary conditions (see also \cite[Chap.~8]{Bergman2011}). Then, the energy equation~\eqref{energy} becomes:
\begin{equation}
\label{energy_ND}
\left(\frac{\rho c_p \mathcal{V}_c H}{q_1} \frac{\partial T}{\partial x} \right) \bar{u}(\bar{y})
= \frac{\partial ^2 \theta}{\partial \bar{y}^2} + 
Br' \left|\frac{d\bar{u}}{d\bar{y}}\right|^{n-1}\left(\frac{d\bar
u}{d\bar{y}}\right)^2,
\end{equation}
where recall that $\mathcal{V}_c$ is the characteristic fluid velocity as in Eq.~\eqref{velocity_scale} above, and $Br'$ is a modified Brinkman number for power-law fluids \cite{BarlettaFullyDissipation}:
\begin{equation}
\label{Br}
Br'=\frac{(\mathcal{V}_c)^{n+1}a}{q_1H^{n}}.
\end{equation}
$Br'$ is a measure of the relative importance of viscous dissipation in the flow compared to the imposed (constant) heat flux. 

Next, note that the quantity in the parentheses on the left-hand side of Eq.~\eqref{energy_ND} is, in fact, dimensionless. Under our assumption of  uniform heat flux being supplied through the upper and lower walls of the microchannel (recall Fig.~\ref{scheme} and the attendant discussion), we shall show that this quantity is, additionally, \emph{constant}. First, let 
\begin{equation}
\label{gamma0}
\gamma = \frac{\rho c_p \mathcal{V}_c H}{q_1}\frac{\partial T}{\partial x}. 
\end{equation}
At this stage in the analysis $\gamma$ is  \emph{unknown} but will be determined shortly. Nevertheless, let us write Eq.~\eqref{energy_ND} as 
\begin{equation}
\label{energy_ND2}
\gamma\bar{u}=\frac{\partial^2\theta}{\partial \bar{y}^2} + Br'\left|\frac{d\bar{u}}{d\bar{y}}\right|^{n-1}\left(\frac{d\bar{u}}{d\bar{y}}\right)^2.
\end{equation}
The pertinent boundary conditions for Eq.~\eqref{energy_ND2} are
\begin{subequations}\begin{alignat}{2}
\theta&=0,\qquad &\bar{y}=0, \label{thermal_BC1}\\
\frac{\partial \theta}{\partial \bar{y}} &=-1,\qquad &\bar{y}=0,\label{thermal_BC2}\\
\frac{\partial \theta}{\partial \bar{y}} &=\frac{q_2}{q_1},\qquad &\bar{y}=1.\label{thermal_BC3}
\end{alignat}\label{thermal_BC}\end{subequations}
Although Eq.~\eqref{energy_ND2} is of second order in $\bar{y}$, we need the \emph{three} boundary conditions in Eqs.~\eqref{thermal_BC} to fully specify the problem because, at this point, $\gamma$ is still unknown.

Now, we are in a position to solve for $\gamma$ by integrating  Eq.~\eqref{energy_ND2} with respect to $\bar{y}$ across the channel height:
\begin{equation}
\gamma \int_{0}^{1}\bar{u}(\bar{y}) \,d\bar{y} =  \int_0^1\frac{\partial^2\theta}{\partial\bar{y}^2} \,d\bar{y} + Br' \int_0^1\left|\frac{d\bar{u}}{d\bar{y}}\right|^{n-1}\left(\frac{d\bar{u}}{d\bar{y}}\right)^2 d\bar{y}.
\end{equation}
Solving for $\gamma$ and using the boundary conditions from Eqs.~\eqref{thermal_BC}, we obtain:
\begin{equation}
\label{gamma}
\gamma = \left[\frac{q_2}{q_1}+1+ Br'\int_0^1\left|\frac{d\bar{u}}{d\bar{y}}\right|^{n-1} \left(\frac{d\bar{u}}{d\bar{y}}\right)^2 d\bar{y}\right] \Big\slash \left[\int_{0}^{1} \bar{u}(\bar{y}) \,d\bar{y}\right] .
\end{equation}
The definite integrals in Eq.~\eqref{gamma} can be evaluated exactly based on the velocity profile given in Eq.~\eqref{velocity_nd}; the resulting expressions are given in \ref{app:integrals}. Now, $\gamma$ is substituted into Eq.~\eqref{energy_ND2}, reducing the latter equation to an ODE in $\theta$, which can be easily solved via standard numerical integration techniques.

\subsection{Nusselt number}
The literature on convective heat transfer and fluid mechanics is full of important dimensionless parameters \cite{Bergman2011}. Amongst the more useful ones in convection heat transfer, and the first one we encounter herein, is the Nusselt number. In fully developed flow, it is customary to utilize the mean temperature $T_m$ as opposed to the centerline temperature for the definition of the Nusselt number. The mean temperature is given by 
\begin{equation}
T_m = \frac{\displaystyle \int_0^H u T \,dy}{\displaystyle \int_0^H u \,dy}.
\end{equation}
Thus, the dimensionless mean temperature is
\begin{equation}\begin{aligned}
\theta_m &=\frac{\displaystyle \int_0^1\bar{u}\theta \,d\bar{y}}{\displaystyle \int_0^1\bar{u}\,d\bar{y}}\\
&=\frac{k}{q_1H}\frac{\displaystyle \int_0^H u \cdot (T-T_{w1})\,dy}{\displaystyle \int_0^H u\,dy}\\
&=\frac{k}{q_1H}(T_m-T_{w1}),
\end{aligned}\end{equation}
which is, again, independent of $x$ based on the foregoing discussion.

Now, the \emph{heat transfer coefficient} ($HTC$) based on the mean temperature of the fluid can be defined as
\begin{equation}
\label{HTC}
HTC = \frac{q_1}{T_{w1}-T_m}.
\end{equation}
Finally, the Nusselt number (for flow between parallel plates) is simply the dimensionless heat transfer coefficient:
\begin{equation}
\begin{aligned}
\label{Nu}
Nu &= \frac{HTC}{k/(2H)}\\
&= \left(\frac{q_1}{T_{w1}-T_m}\right)\frac{2H}{k}\\
&=-\frac{1}{\theta_m}.
\end{aligned}
\end{equation}
Using the temperature profile obtained by numerically integrating the ODE \eqref{energy_ND2} and the analytical velocity profile from Eq.~\eqref{velocity_nd}, $Nu$ can be evaluated numerically directly from Eq.~\eqref{Nu}.

\subsection{Entropy generation}

The expression for volumetric entropy generation due to convection heat transfer with viscous dissipation included (see, e.g., \cite{Anand2014}) in a power-law fluid is given by
\begin{equation}
\label{Sdotppp}
\dot{S}_\mathrm{gen} = \frac{k}{T^2}\left[\left(\frac{\partial T}{\partial x}\right)^2 +\left(\frac{\partial T}{\partial y}\right)^2\right ]+\frac{a}{T}\left|\frac{du}{dy}\right|^{n+1}.
\end{equation}
The dimensionless form of Eq.~\eqref{Sdotppp} is
\begin{equation}\label{Sdotppp_ND}\begin{aligned}
N_s &= \frac{\dot{S}_\mathrm{gen}}{k/H^2} \\
&=\frac{\phi^2}{(\phi\theta+1)^2}\frac{\gamma^2}{Pe^2}+\frac{\phi^2}{(\phi\theta+1)^2}\left(\frac{\partial\theta}{\partial\bar{y}}\right)^2 \\
&\phantom{=} +\frac{Br'}{\phi\theta+1}\phi  \left|\frac{d\bar{u}}{d\bar{y}}\right|^{n+1}.
\end{aligned}\end{equation}
Here, 
\begin{equation}
\phi=\frac{q_1}{Hk}
\end{equation}
is the dimensionless heat flux, and
\begin{equation}
Pe=\frac{\rho c_p \mathcal{V}_c H}{k}
\end{equation}
is the P\'eclet number. 

We are interested in the limit $Pe\gg1$, hence the first term in Eq.~\eqref{Sdotppp} can be neglected, consistent with our assumption of a fully developed thermal field. Then, we obtain the following simplified expression for the dimensionless entropy generation:
\begin{equation}
\label{N_s}
N_s(\bar{y}) = \frac{\phi^2}{(\phi\theta+1)^2}\left(\frac{\partial\theta}{\partial\bar{y}}\right)^2+\frac{Br'}{\phi\theta+1}\phi  \left|\frac{d\bar{u}}{d\bar{y}}\right|^{n+1}.
\end{equation}
The first term on the right-hand side of Eq.~\eqref{N_s} stands for entropy generation due to heat transfer, while the second term stands for entropy generation due to fluid flow.

The entropy generation rate $N_s$, defined in Eq.~\eqref{Average} for our flow, is a function of the cross-sectional coordinate $\bar{y}$ and, thus, will vary across the cross-section. However, unlike the temperature and velocity distributions, but like the flow rate and the total energy, entropy can also be expressed as an \emph{extensive} property. Thus, it is more insightful to study its behavior on an \emph{average} basis, as opposed to studying its behavior as a field, i.e., a distribution, in $\bar{y}$. To this end, we define the average (dimensionless) entropy generation rate across the channel height:
\begin{equation}
\label{Average}
\langle N_s\rangle =\int_0^1 N_s \,d\bar{y}.
\end{equation}

\subsection{Bejan number}
The expression for entropy generation rate does not convey which of the two entropy generation mechanisms---fluid flow and heat transfer---dominates. Another dimensionless number, termed the \emph{Bejan number}, can be defined to ascertain which mechanism of entropy generation is more significant \cite{paoletti1989calculation} (see also the discussion in \cite{Petrescu1994}). The Bejan number $Be$ is the ratio of entropy generated due to heat transfer to the total entropy generated:
\begin{equation}
\label{Be}
Be(\bar{y}) = \frac{\displaystyle \frac{\phi^2}{(\phi\theta+1)^2}\left(\frac{\partial\theta}{\partial\bar{y}}\right)^2}{N_s}.
\end{equation}
Notice that, just as $N_s$ in Eq.~\eqref{N_s}, $Be$ varies across the channel width, suggesting that the relative proportion of entropy generation due to heat transfer is different at different vertical locations in the channel.

\section{Results and discussion}
\label{sec:results}

The equations for momentum transfer, heat transfer and entropy generation have been derived, solved exactly or reduced to quadratures in Sec.~\ref{S:2}. In this section, we would like to discuss  the fluid mechanics, heat transfer, and thermodynamics of power-law fluids in heated microchannels with asymmetric slip for particular/illustrative values of the parameters, as tabulated in Table~\ref{tb:params}. In doing so, we present some novel aspects to the physics of this system.

\begin{table}[ht]
\small
\centering
\begin{tabular}{l l l}
\hline
\textbf{Parameter} & \textbf{Symbol} & \textbf{Value(s)}\\
\hline
Brinkman number & $Br'$ & 0.2 \\
Dim'less slip coeff., wall 1 & $\bar{K}_1$ & 0.5, 0.75, 1.0, 1.25 \\
Dim'less slip coeff., wall 2 & $\bar{K}_2$  & 1.0 \\
Slip exponent & $m$ & 0.7\\
Power-law index & $n$ & 0.5, 1.5\\
Ratio of heat fluxes & $q_2/q_1$ & 0.5, 2.0\\
Dimensionless heat flux & $\phi$ & $0.4$ \\
\hline
\end{tabular}
\caption{Numerical values assigned/considered in our analysis for the various parameters arising in the model under consideration.}
\label{tb:params}
\end{table}

Note that we have kept $Br'$ constant. The reason is that the effects of $Br'$ on heat transfer and entropy generation characteristics of power-law fluid have already been addressed in detail by Tso et al.~\cite{Tso2010} and Hung \cite{Hung2008}. The value of power-law index $n$ has been taken to be either 0.5 or 1.5 to cover both the shear-thinning and shear-thickening aspects of non-Newtonian fluids.  The dimensionless slip coefficient on upper wall is constant, $\bar{K}_2=1$, while the corresponding one on the lower wall, $\bar{K}_1$, is given four different values.  This is in agreement with the theme of the paper, which is to study the effect of such asymmetric slip on the posed thermofluids problem.  We have also allowed asymmetry in imposed uniform heat fluxes on the channel walls through the ratio $q_2/q_1$.

The results in the figures discussed below were generated by first solving numerically the non-linear algebraic Eq.~\eqref{velocity_nd_tau_eq} for $\bar{\tau}_{w1}$ given $\bar{K}_1$, $\bar{K}_2=1$, $n$ and $m$. Then, this value of $\bar{\tau}_{w1}$, together with Eqs.~\eqref{velocity_nd} and \eqref{velocity_nd_bc}, fully specifies the velocity profile. The numerical solution is obtained using standard subroutines from SciPy \cite{SciPy}. The dimensionless velocity profile thus obtained is used to determine the dimensionless temperature profile across the channel by solving Eq.~\eqref{energy_ND2} numerically, after first evaluating $\gamma$ via Eq.~\eqref{gamma}. To this end, we use the interleaved 4(5) Dormand--Prince pair explicit Runge--Kutta method, namely the scipy.integrate.ode function {\tt dopri5}, with a relative error tolerance of $10^{-3}$. The Nusselt number is obtained by substituting the dimensionless mean temperature into Eq.~\eqref{Nu}. The dimensionless velocity and temperature profiles are finally also substituted into Eqs.~\eqref{Average} and \eqref{Be}, and numerical integration is used to obtain the average entropy generation rate $\langle N_s\rangle$ and the Bejan number distribution $Be(\bar{y})$.

\begin{figure}
\centering\includegraphics[width=\linewidth]{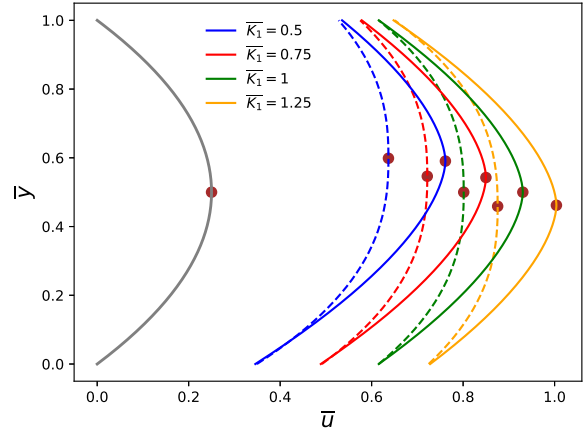}
\caption{(Color online.) Dimensionless velocity profiles $\bar{u}(\bar{y})$ across the channel for different values of the dimensionless slip coefficient $\bar{K}_1$. Dashed curves correspond to $n=0.5$ (shear-thinning fluid), while solid curves correspond to $n=1.5$ (shear-thickening fluid). The gray curve on the left corresponds to a Newtonian fluid subject to no slip boundary conditions. Filled circles denote the vertical location of the velocity maximum.}
\label{scheme2}
\end{figure}

\subsection{Velocity profile}
\label{VP}

\begin{figure}
\centering\includegraphics[width=\linewidth]{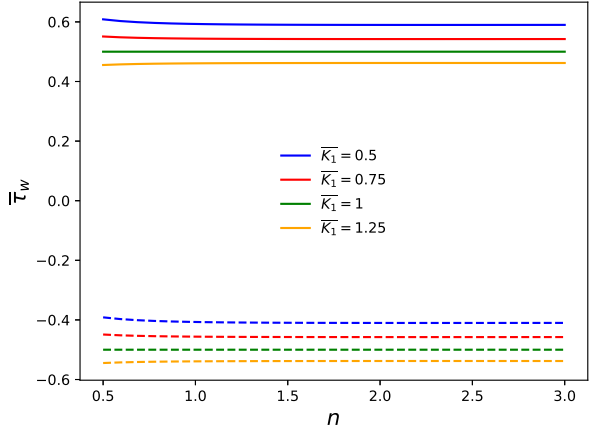}
\caption{(Color online.) Dimensionless shear stress at the channel walls $\bar{\tau}_w$ versus the power-law index $n$ for different values of the slip coefficient $\bar{K}_1$. The dashed lines show the shear stress on the upper wall, $\bar{\tau}_{w2}$, while the solid curves show the shear stress on the lower wall, $\bar{\tau}_{w1}$. }
\label{scheme3}
\end{figure}

Figure~\ref{scheme2} shows  the velocity profile across the channel, as determined by  Eqs.~\eqref{velocity_nd}--\eqref{velocity_nd_tau_eq}, for different values of the slip coefficient $\bar{K}_1$. The velocity profiles for shear-thinning ($n<1$) and shear-thickening ($n>1$) fluids are shown by the dashed and solid curves in the plot, respectively. This convention is followed throughout this paper with the exception of Fig.~\ref{scheme3}. For the sake of comparison, the velocity profile of a Newtonian fluid satisfying no slip at the walls is also shown in gray. The vertical location at which the velocity gradient changes sign, which is the point of zero shear stress in the channel, is denoted by a filled circle on each curve.

As can be seen immediately from the graph, the presence of slip increases the maximum velocity of  both shear-shinning and shear-thickening fluids. However,  shear-thinning fluids ($n<1$) (dashed curves) exhibit  significantly less steep velocity profiles than shear-thickening fluids ($n>1$) (solid curves). This observation can be explained by noting that shear-thinning fluids will support a smaller shear stress and, consequently by Eq.~\eqref{power_law}, smaller velocity gradients.

We also see that, for a fixed value of $n$, the velocity profile \emph{and} its gradient are \emph{both} affected by the asymmetric slip coefficient's magnitude. This is a new result due to asymmetric slip because, as shown in  \cite{Anand2014}, the velocity profile of a power-law fluid subject to symmetric slip will have the same slope for all values of the slip coefficient, which is also well known for Newtonian microflow \cite[Sect.~24.4]{Panton}. Therefore, for symmetric slip boundary conditions, slip affects only the advection of momentum but not its diffusion. Specifically, if $\bar{u}_0(\bar{y})$ is the solution for the velocity profile without slip, then the velocity profile with symmetric slip can always be written as $\bar{u}(\bar{y}) = \bar{u}_0(\bar{y}) + \hat{u}$, where $\hat{u}$ is some constant. In such a symmetric slip flow, one could thus model the problem simply as a shear-driven Couette flow, i.e., as the flow in a channel with walls moving at some appropriate velocities, as done in \cite{Shojaeian2014ConvectiveConditions}. \emph{However}, for the case of asymmetric slip treated herein, slip affects both the diffusion and advection of fluid momentum. Thus, a flow with asymmetric slip \emph{cannot} simply be modeled as a Couette flow with moving walls, which should also be clear from the derivation in Sect.~\ref{sec:vel}, specifically the form of the solution given in Eq.~\eqref{velocity_nd}.

We also observe that the vertical location of the maximum velocity, $\bar{y}_0\equiv y_0/H = \bar{\tau}_{w1}$, depends on the relative values of the slip coefficients (as parametrized by $\bar{K}_1$ in the present work). But, $\bar{\tau}_{w1}$  \textit{does not depend on the shear-thinning or shear-thickening aspect of the fluid}, as made clear by the lack of $n$ in Eq.~\eqref{velocity_nd_tau_eq}. For the cases of equal slip coefficients (labeled as $\bar{K}_1=1$ in Fig.~\ref{scheme2}), the velocity reaches a maximum at mid-height of the channel, i.e., at $\bar{y}_0 = \bar{\tau}_{w1} = 1/2$, as mentioned in the discussion following Eq.~\eqref{velocity_nd_tau_eq}. However, for the case of asymmetric slip, the location of the maximum is closer to the wall on which the slip is larger (the upper wall for blue and red curves and the lower wall for yellow curves). Additionally, the velocity gradients are larger near the wall on which $\bar{K}$ is smaller. To understand this aspect, note that slip reduces friction on the walls. Consequently, the shear stress (and the velocity gradient) will be smaller on the wall where the slip is larger. Thus, the point of maximum velocity (i.e., zero velocity gradient) must be closer to the wall with larger $\bar{K}$.

To buttress the above point, in Fig.~\ref{scheme3} we plot the shear stress at the walls $\bar{\tau}_{w\{1,2\}}$ (as solid and dashed curves, respectively) as a function of the power-law index $n$ for different values of the slip coefficient $\bar{K}_1$. Due to the choice of coordinate system, $\bar{\tau}_w$ is positive on the lower wall (solid curves) but negative on the upper wall (dashed curves). Consequently, Fig.~\ref{scheme3} shows that enhanced slip reduces the shear stress at the wall since $\bar{\tau}_{w1}$ decreases with $\bar{K}_1$. However, a decrease in $\bar{\tau}_{w1}$ is accompanied by an increase in the (absolute value of the) shear stress at the upper wall, $\bar{\tau}_{w2}$, even though the slip coefficient at the upper wall is constant. This observation is explained by noting that $\bar{\tau}_{w2} = \bar{\tau}_{w1}-1$ ($0<\bar{\tau}_{w1}<1$), after making Eq.~\eqref{stress_tw} dimensionless, thus the two wall shear stresses are coupled. 

\begin{figure}
\centering\includegraphics[width=\linewidth]{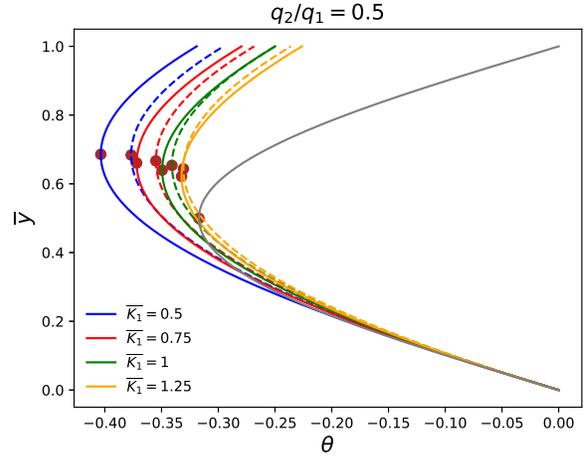}
\caption{(Color online.) Dimensionless temperature distribution $\theta$ across the height $\bar{y}$ of the channel with a larger imposed heat flux at the lower wall ($q_2/q_1=0.5 <1 $), for different values of the slip coefficient $\bar{K}_1$. The dashed curves show the profiles of a shear-thinning fluid with power-law exponent $n=0.5$, while the solid curves show the profiles of a shear-thickening fluid with $n=1.5$.}
\label{scheme4}
\end{figure}

\subsection{Temperature distribution}
\label{TD}

There are three modes of heat transfer in the problem that we posed in Sect.~\ref{S:1}, namely: axial advection, vertical conduction, and viscous dissipation. Axial advection is represented by the term on the left-hand side of Eq.~\eqref{energy}, vertical conduction is represented by the first term on its right-hand side, while viscous dissipation is represented by the second term on the right-hand side of Eq.~\eqref{energy}. Consequently, the temperature distribution is intimately coupled to the velocity field through the axial advection and viscous dissipation terms. Clearly, the fluid flow problem needs to be solved prior to attacking the heat transfer (and entropy generation) problem. This ordering is also the reflected in the history of convection heat transfer research: the Blasius solution for the laminar hydrodynamic boundary layer \cite{Blasius} preceded the Pohlhausen solution for the corresponding thermal boundary layer \cite{Pohlhausen} by thirteen years (see also the discussion in \cite{Jessee2015}). 

\begin{figure}
\centering\includegraphics[width=\linewidth]{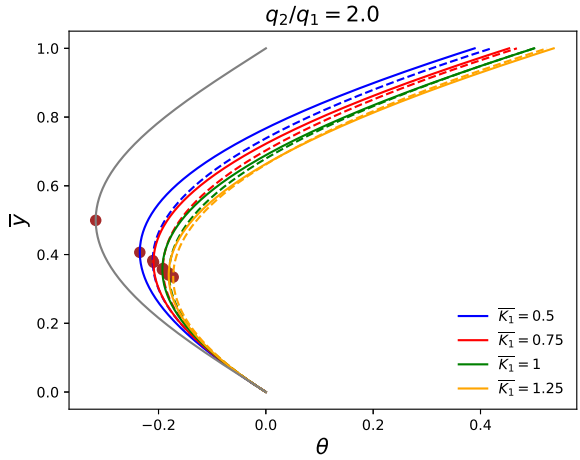}
\caption{(Color online.) Dimensionless temperature distribution $\theta$ across the height $\bar{y}$ of  the channel with a larger imposed heat flux at the upper wall ($q_2/q_1=2.0>1$), for different values of the slip coefficient $\bar{K}_1$. The dashed curves show the profiles of a shear-thinning fluid with power-law exponent $n=0.5$, while the solid curves show the profiles of a shear-thickening fluid with $n=1.5$.}
\label{scheme5}
\end{figure}

Likewise, we solved the fluid mechanical problem in Sect.~\ref{sec:vel} and the temperature distribution was analyzed in Sect.~\ref{sec:temp}. In this subsection, we first  analyze the temperature profile across the channel height for different values of the slip coefficient $\bar{K}_1$ in Figs.~\ref{scheme4} and \ref{scheme5}, for $q_2/q_1 =0.5$ and $q_2/q_1 =2.0$, respectively. To highlight the physics studied herein, the temperature profile corresponding to a Newtonian fluid subject to \emph{no} slip ($\bar{K}_1=\bar{K}_2=0$) and with equal applied heat fluxes ($q_2/q_1=1$) on the walls is shown as a gray curve. Observe that, from Eq.~\eqref{theta},  $\theta=0$ at $\bar{y}=0$ while the imposed wall heat flux is into the fluid, so the temperature in the interior must be lower than that of wall ``1,'' resulting in negative values of $\theta$ as depicted in Figs.~\ref{scheme4} and \ref{scheme5}.

Next, we observe that the overall temperature distribution inside of the channel is damped out (becomes less negative) with increasing $\bar{K}_1$. This nontrivial effect can be rationalized by noting that an increase in slip leads to an increase in the fluid velocity and a concomitant increase in the ability of the fluid to carry heat, resulting in higher temperatures. Thus, for a fixed heat flux ratio $q_2/q_1$, we see that the minimum of $\theta$ decreases with $\bar{K}_1$. In {both} cases ($q_2/q_1=0.5$ and $q_2/q_1=2.0$), the  vertical $\bar{y}$-location of the minimum (i.e., where $\partial\theta/\partial \bar{y}=0$) moves closer to the \emph{lower} wall as the slip coefficient increases.

Now, comparing Fig.~\ref{scheme4} to Fig.~\ref{scheme5}, we further understand that the location at which $\partial\theta/\partial \bar{y}=0$ is closer to the cooler wall (specifically, the upper wall in Fig.~\ref{scheme4} and the lower wall in Fig.~\ref{scheme5}), where the heat flux is smaller. This effect is attributed to the fact that a higher heat flux at the wall drives a higher temperature gradient into the fluid, owing to Fourier's law of heat conduction. 

The final observation that we would like to make about Figs.~\ref{scheme4} and \ref{scheme5} is that the effect of increasing the slip coefficient is more pronounced when the lower wall is hotter (i.e., $q_2/q_1 < 1$). This is because the two phenomena discussed above---higher heat flux and larger slip at the lower wall---now work in tandem to increase the temperature at the lower wall. Conceivably, this synergy between asymmetric slip and asymmetric imposed heat flux could be exploited in microscale heat transfer applications.

\begin{figure}
\centering\includegraphics[width=\linewidth]{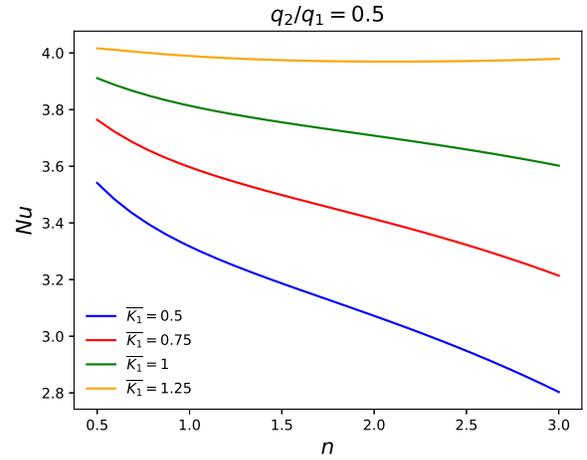}
\caption{(Color online.) The Nusselt number along the lower wall ($\bar{y}=0$) for different values of the power-law exponent $n$ and the slip coefficient $\bar{K}_1$. The ratio of imposed heat fluxes is ${q_2}/{q_1}=0.5$. }
\label{scheme6}
\end{figure}

\subsection{Nusselt number}

Recall that the Nusselt number is the dimensionless heat transfer coefficient ($HTC$), as given by Eq.~\eqref{Nu}, which is the ratio of the heat flux to the temperature difference between the fluid and the lower channel wall. Although a corollary of Newton's law of cooling, as discussed by Bejan~\cite{Bejan} the concept of a heat transfer coefficient is, in fact, inherently ambiguous. For example, in natural convection the heat transfer coefficient is necessarily a function of the temperature difference and is, thus, \emph{not} constant \cite{Bergman2011}. On the other hand, the widespread use of computational fluid dynamics (CFD), wherein the energy balance at the surface is enforced explicitly by mapping the temperatures or heat fluxes directly from one domain to another \cite{Illingworth}, has allowed for more general analyses of thermofluids problems, often obviating the need for the Nusselt number as a fundamental physical quantity. We also understand that the Nusselt number is completely defined only by specifying a reference temperature, see Eq.~\eqref{Nu}, which itself depends on the rate of heat transfer---a quantity that we are trying to predict in the first place! This makes the definition of the Nusselt number a circular one. 

Nevertheless, for forced convection in unidirectional flows, as in the present study, the Nusselt number remains a relevant quantity that can be used to understand the thermal characteristics of heat transfer in the flow. Indeed, this has been the quantity of interest in a large number of scientific works on heat transfer in microchannel flows \cite{gs03,Yovanovich2015,Tso2010,Sheela-Francisca2012,Anand2014,Shojaeian2014ConvectiveConditions}.

\begin{figure}
\centering\includegraphics[width=\linewidth]{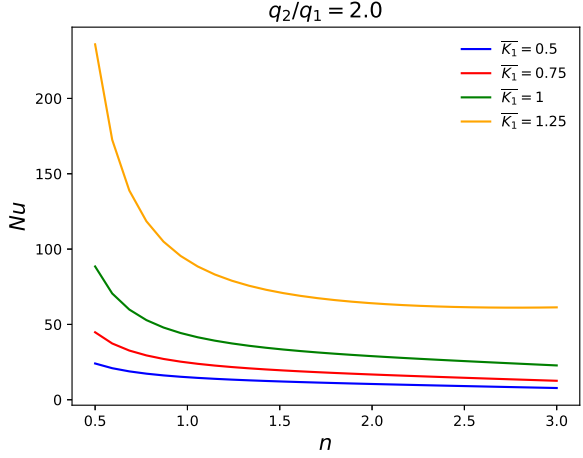}
\caption{(Color online.) The Nusselt number along the lower wall ($\bar{y}=0$) for different values of the power-law exponent $n$ and the slip coefficient $\bar{K}_1$. The ratio of imposed heat fluxes is ${q_2}/{q_1}=2.0$.}
\label{scheme7}
\end{figure}

In the present work, we have imposed uniform heat fluxes (independent of the axial coordinate, $x$) on the channel walls. However, we have allowed the uniform heat flux on the lower wall, $q_1$, to be different from the uniform heat flux on the upper wall, $q_2$. Consequently, $Nu$ along the lower wall ($\bar{y}=0$) necessarily differs from $Nu$ along the upper wall ($\bar{y}=1$). Still, $Nu$ is independent of $x$. This result has been shown with due mathematical rigor by Tso et. al.~\cite{Tso2010}. Thus, in Figs.~\ref{scheme6} and \ref{scheme7}, we only plot the Nusselt number $Nu$ along the lower wall as a function of the power-law index $n$ of the fluid and for different values of the slip coefficient $\bar{K}_1$. Specifically, Fig.~\ref{scheme6} shows the $Nu$ dependence on $n$ for a heat flux ratio $q_2/q_1=0.5$, while Fig.~\ref{scheme7} shows the $Nu$ dependence on $n$ for a heat flux ratio $q_2/q_1=2.0$.

From Figs.~\ref{scheme6} and \ref{scheme7}, we immediately conclude that the Nusselt number increases with the amount of slip. From our earlier discussion of the temperature profile: an increase in the wall slip increases the fluid's ability to advect heat near the walls, leading to enhanced heat transfer. On the other hand, $Nu$ decreases with the power-law index $n$, consistent with previous studies \cite{Shojaeian2014ConvectiveConditions,Anand2014}. For small values of $n$, the shear-thinning aspect of the fluid is more prominent, reducing resistance to flow and increasing the velocity. Consequently, the flow's ability to transport heat is enhanced.

Finally, we see that $Nu$ increases \emph{significantly} with the ratio $q_2/q_1$ (as can be obviously inferred by comparing the vertical axes scales between Figs.~\ref{scheme6} and \ref{scheme7}). On one hand, this dependence is in stark contrast to the case of equal imposed heat fluxes (when $q_2/q_1=1$), in which the Nusselt number is independent of the value of $q_2=q_1$ \cite{Anand2014}. On the other hand, this dependence is consonant with previous studies with unequal imposed heat fluxes in which $Nu$ depends on $q_2/q_1$ \cite{Tso2010,Sheela-Francisca2012}.\footnote{As a consistency check, note that the Nusselt number for the \emph{special case} of a Newtonian fluid subject to no slip and equal imposed heat fluxes was found to be 4.1176 based on our mathematical results, which agrees with the value reported in literature \cite{Tso2010} to four decimals.} This again suggests that such an enhancement of heat transfer due to asymmetric slip and asymmetric imposed heat fluxes could be exploited in the design of microscale heat transfer equipment.

\subsection{Average entropy generation}

\begin{figure}
\centering\includegraphics[width=\linewidth]{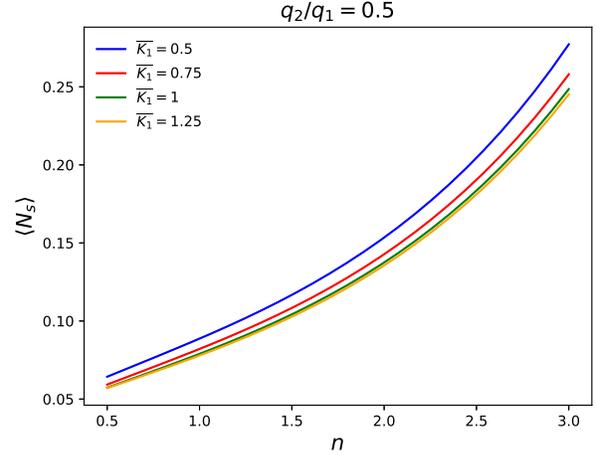}
\caption{(Color online.) Average entropy generation rate $\langle N_s\rangle$ as a function of the power-law index $n$ for different values of the slip coefficient $\bar{K}_1$. The ratio of imposed heat fluxes is ${q_2}/{q_1}=0.5$.}
\label{scheme8}
\end{figure}

\begin{figure}
\centering\includegraphics[width=\linewidth]{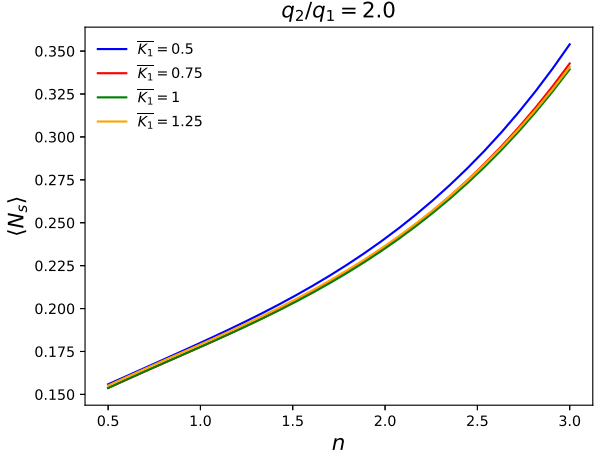}
\caption{(Color online.) Average entropy generation rate $\langle N_s\rangle$ as a function of the power-law index $n$ for different values of the slip coefficient $\bar{K}_1$. The ratio of imposed heat fluxes is ${q_2}/{q_1}=2.0$.}
\label{scheme9}
\end{figure}

We have defined the \emph{average} entropy generation rate $\langle N_s\rangle$ in Eq.~\eqref{Average}. In this subsection, the average entropy generation rate has been plotted as a function of the power-law index $n$ of the fluid in Figs.~\ref{scheme8} and \ref{scheme9} for $q_2/q_1=0.5$ and $q_2/q_1=2.0$, respectively. 

First, it is seen from Figs.~\ref{scheme8} and \ref{scheme9} that $\langle N_s\rangle$ increases with the power-law index $n$. This observation is attributed to the fact that an increase in $n$ leads to higher velocity and temperature gradients inside the channel, as already explained in Sections~\ref{VP} and \ref{TD}. Consequently, more entropy is generated by both heat transfer and viscous friction.

Second, it is evident from Figs.~\ref{scheme8} and \ref{scheme9} that $\langle N_s\rangle$  decreases with the slip coefficient at the walls. As we have already seen in Sections~\ref{VP} and \ref{TD}, a larger slip leads to more thorough mixing and more uniform temperature and velocity distributions inside the flow. Thus, the inherent thermodynamic irreversibility of the flow (as measured by the entropy generation rate) is diminished as slip increases.

Lastly, $\langle N_s\rangle$ increases with the ratio of imposed heat fluxes, $q_2/q_1$. This trend is  expected because imposing larger heat fluxes leads to larger temperature gradients in the channel, and hence to further thermodynamic irreversibility.

\subsection{Bejan number}

In thermal engineering, researchers concern themselves with more than just the amount of entropy generated. Their imperative is to make the system more efficient by minimizing the entropy generated inside it \cite{BejanAdr1999}. To that end, they need to understand the various modes by which entropy is generated inside the physical system under investigation. The question ``How is entropy minimized?'' then begets the question ``How is entropy generated?''. Paoletti et al.~\cite{paoletti1989calculation} defined the Bejan number to address this issue.

The Bejan number $Be$ is the ratio of the entropy generated due to heat transfer to the total entropy generated, as given in Eq.~\eqref{Be}. It follows that an increase in fluid flow irreversibility (velocity gradients) causes the Bejan number to decrease, while an increase in heat transfer irreversibility (temperature gradients) causes the Bejan number to increase. By this definition, it is obvious that for $\bar{y}$ such that $Be(\bar{y}) >1/2$, the entropy generation is primarily due to heat transfer irreversibility, and conversely, fluid friction irreversibility dominates in the region(s) where $Be(\bar{y}) <1/2$. The two entropy generation mechanisms are of equal ``importance'' when/where $Be=1/2$.

\begin{figure}
\centering\includegraphics[width=\linewidth]{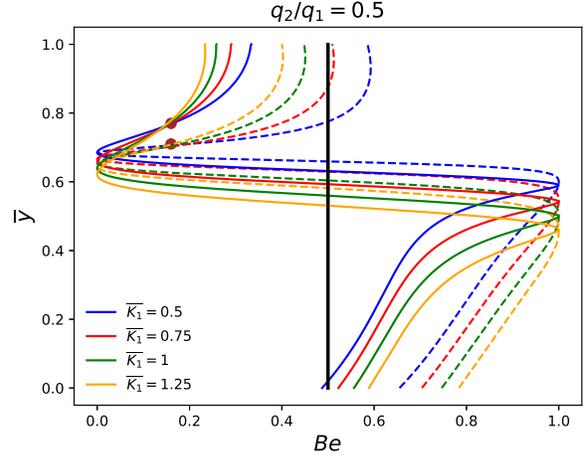}
\caption{(Color online.) Distribution of Bejan number $Be(\bar{y})$ across the channel width for different values of the slip coefficient $\bar{K}_1$ and $q_2/q_1=0.5$. The dashed curves correspond to a shear-thinning fluid with $n=0.5$, while the solid curves correspond to a shear-thickening fluid with $n=1.5$. The black vertical denotes $Be=1/2$.}
\label{scheme10}
\end{figure}

\begin{figure}
\centering\includegraphics[width=\linewidth]{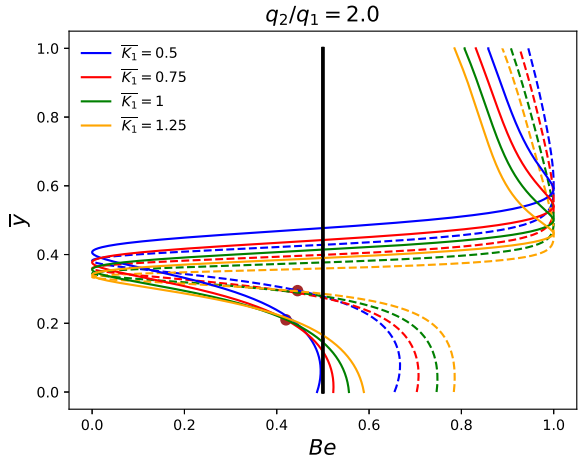}
\caption{(Color online.) Distribution of Bejan number $Be(\bar{y})$ across the channel width for different values of the slip coefficient $\bar{K}_1$ and $q_2/q_1=2.0$. The dashed curves correspond to a shear-thinning fluid with $n=0.5$, while the solid curves correspond to a shear-thickening fluid with $n=1.5$. The black vertical denotes $Be=1/2$.}
\label{scheme11}
\end{figure}

The distribution of the Bejan number $Be(\bar{y})$ across the channel width for different slip coefficientsis shown in Figs.~\ref{scheme10} and \ref{scheme11} for $q_2/q_1=0.5$ and $q_2/q_1=2.0$, respectively. The global maximum (here, at $Be=1$ for all cases shown) of each curve corresponds to the point at which the velocity gradient vanishes ($\partial\bar{u}/\partial\bar{y}=0$), while the global minimum (here, at $Be=0$ for all cases shown)  corresponds to the point at which the temperature gradient vanishes ($\partial\theta/\partial\bar{y}=0$). For a flow with symmetric slip ($\bar{K}_1=1$) and equal heat flux boundary conditions ($q_2/q_1=1$), both the temperature and velocity gradients vanish exactly at the centerline ($\bar{y}=1/2$), making it a point of singularity for the $Be$ profile \cite{Anand2014}. The vertical line $Be =1/2$ denotes the line of demarcation for a trade-off in the relative importance of the two entropy generation mechanisms. We observe from both Figs.~\ref{scheme10} and \ref{scheme11} that a particular trade-off point exists near the centerline of the channel (between the global maximum and the global minimum of $Be$ heretofore described) for both shear-thinning and shear-thickening fluids and for all values of the slip coefficient, \emph{unlike the case of symmetric slip} \cite{Anand2014}.

For the same value of $q_2/q_1$, i.e., restricting to either Fig.~\ref{scheme10} or Fig.~\ref{scheme11}, it is seen that the global maximum of the $Be$ curve shifts towards the lower wall when the slip coefficient on the lower wall, namely $\bar{K}_1$, increases. This trend is explained by recalling the velocity profiles in Fig.~\ref{scheme2}, in which the location of the maximum of the velocity profile shifts towards the lower wall as $\bar{K}_1$ increases. 

In both Figs.~\ref{scheme10} and \ref{scheme11}, we see that the behavior of the $Be$ curves is different at each wall. At the lower wall, the Bejan number is smaller for smaller slip, i.e., the blue curve (smaller $\bar{K}_1$) is to the left of the yellow curve (larger $\bar{K}_1$). However, the situation is reversed where the curves terminate on the upper wall. The spanwise location at which this reversal or ``switch'' occurs, i.e., the $\bar{y}$ value at which the curves cross each other, is shown with a filled circle in each figure. This circle is closer to the cooler wall in each of Figs.~\ref{scheme10} and \ref{scheme11}. We understand from our discussion of the velocity distribution in Sect.~\ref{VP} that as the slip coefficient $\bar{K}_1$ increases, the shear stress $\bar{\tau}_{w1}$ on the lower wall decreases; this leads to an increase in $Be$ near said wall. However, the reverse holds true at the upper wall, where an increase in $\bar{K}_1$ leads to a corresponding increase in $\bar{\tau}_{w2}$ and a decrease in $Be$. The location at which this reversal occurs is closer to the cooler wall because the temperature gradients there are smaller, i.e., the relative influence of velocity gradients is more pronounced.

We also see that, overall, $Be(\bar{y};n<1)>Be(\bar{y};n>1)$, i.e., the Bejan number for shear-thinning fluids is higher than that for shear-thickening fluids, in most parts of the microchannel. To explain this observation, note that shear-thinning fluids exhibit smaller velocity gradients compared to shear-thickening fluids (i.e., less resistance to flow), which translates means that a larger portion of the entropy generation is due to heat transfer.

Next, we observe that in Figs.~\ref{scheme10} and \ref{scheme11} there is a \emph{second} (local) maximum in the $Be(\bar{y})$ profile near the cooler wall for shear-thinning fluids for all values of slip coefficients (and also for shear-thickening fluids with $\bar{K}_1<1$). This second (local) maximum is absent in the case of symmetric slip \cite{Anand2014}.  To understand this observation, note that the velocity gradients in a shear-thinning fluid quickly diminish away from the walls, while the temperature gradients do not drop-off so steeply (recall the discussion in Sects.~\ref{VP} and \ref{TD}). This discrepancy in how fast velocity and temperature gradients are dissipated in this microflow leads to the presence of the local maximum in the $Be$ profile. 

Finally, we note that some of the curves corresponding to shear-thinning fluids with $\bar{K}_1>1$ in Fig.~\ref{scheme10}, and all the curves in Fig.~\ref{scheme11}, cross the vertical line $Be = 1/2$ \emph{twice}: near the channel centerline (as discussed previously) and again close to the cooler wall.\footnote{To be precise, however, we also note that, in Fig.~\ref{scheme10}, some of the shear-thickening curves with $\bar{K}<1$ appear to approach $Be=1/2$ again near the \emph{hotter} wall, suggesting a highly nontrivial dependence on the model parameters.} Thus the trade-off between the relative importance of the two entropy generation mechanisms happens at two locations for these specific flows. This novel effect also appears to be entirely due to \emph{asymmetric slip}. Evidently, non-Newtonian rheology, asymmetric slip and asymmetric imposed heat fluxes can be used to \emph{tune} how entropy is generated across the channel in a microflow.

\section{Conclusion}

We studied the effect of asymmetric slip on the hydrodynamics, heat transfer and entropy generation of a thermally developed flow of a non-Newtonian fluid in a microchannel subject to asymmetric nonlinear slip and unequal heat fluxes at the walls. Under a power-law rheological model, the governing equations were specialized to viscous unidirectional flow, yielding an exact but implicit solution for the velocity profile. The heat transfer and entropy generation characteristics (specifically,  Nusselt number, average entropy generation rate and Bejan number) were reduced quadratures. 

In summary, we have established the following. First, slip \emph{reduces} the shear stress acting on the channel walls, especially at the wall with ``larger'' slip. Second, slip \emph{promotes} uniformity in the temperature field. The temperature in the fluid  increases with slip asymmetry, in particular because of its enhanced capacity to transfer heat by advection. Third, flows with larger slip induce less entropy generation, specifically the Bejan number $Be$ is larger near the hotter wall in the case of unequal imposed heat fluxes. Overall, slip reduces both heat transfer irreversibility and fluid flow irreversibility, however, it has a more pronounced effect on fluid flow irreversibility by directly influencing the velocity profile. Unlike the case of symmetric slip, we found that, for certain combinations of the power-law index $n$, the dimensionless slip asymmetry parameter $\bar{K}_1$, and the imposed heat flux asymmetry $q_2/q_1$ at the channel walls, a second trade-off point between the entropy generation mechanisms exists, apart from the trade-off point closer to the channel's centerline.
  
We believe that our theoretical analysis will be useful in thermal engineering. Specifically, our conclusions could be used to tailor heat transfer in microchannel flows of complex fluids (such as blood or dense polymeric suspensions) to minimize thermal irreversibility, for example, by tuning the channel's surface properties so that the complex fluid's flow exhibits asymmetric slip. Likewise, in the case of unequal boundary heat fluxes, imposing the higher flux on the wall with larger slip leads to enhanced heat transfer into the fluid.

In future work, heat transfer and entropy generation in the asymmetric slip microflows of gases could be attacked along on the lines of \cite{AH17,HA10,SL61}. In such flows, a temperature jump boundary condition emerges at the boundaries, as originally discussed by Smoluchowski \cite{S98}. However, as noted in \cite{STC17}, it is currently an open question whether the temperature jump boundary conditions derived for slip flow of gasses can apply to complex liquids (such as the example dense suspension, e.g., \cite{Vayssade2014,Georgi1}, considered in this work).

\section*{Acknowledgements}
This research was, in part, supported by the US National Science Foundation under grant No.\ CBET-1705637. We thank the reviewers for their helpful comments on the manuscript.

\begin{nomenclature}
\entry{$a$}{consistency index, N$\cdot$s$^{n+1}/$m$^{n+1}$}
\entry{$Br'$}{modified Brinkman number}
\entry{$C$, $C_1$, $C_2$}{constants of integration}
\entry{$c_p$}{specific heat capacity, kJ$/$(kg$\cdot$K)}
\entry{$G$}{pressure gradient, Pa$/$m}
\entry{$H$}{height of the channel, $m$}
\entry{$HTC$}{heat transfer coefficient, W$/$(m$^2\cdot$K)}
\entry{$K$}{slip coefficient, m$\cdot$Pa$^{-m}/$s}
\entry{$k$}{thermal conductivity, W$/$(m$\cdot$K)}
\entry{$\ell_c$}{characteristic length scale, m}
\entry{$m$}{slip exponent}
\entry{$n$}{power-law index}
\entry{$N_s$}{entropy generation number}
\entry{$Nu$}{Nusselt number}
\entry{$Pe$}{P\'eclet number}
\entry{$q$}{imposed heat flux along a wall, W$/$m$^2$}
\entry{$\dot{S}_\mathrm{gen}$}{volumetric entropy generation rate W$/$m$^3\cdot$K}
\entry{$T$}{temperature, K}
\entry{$\mathcal{T}_{c}$}{characteristic stress scale, Pa}
\entry{$u$}{velocity, m/s}
\entry{$\mathcal{V}_c$}{characteristic aial velocity scale, m$/$s}
\entry{$x$}{axial coordinate, m }
\entry{$y$}{transverse/spanwise coordinate, m}
\textbf{Greek Symbols}

\entry{$\tau$}{shear stress, Pa}
\entry{$\gamma$}{dimensionless constant}
\entry{$\rho$}{density, kg$/$m$^3$}
\entry{$\theta$}{dimensionless temperature}
\entry{$\phi$}{dimensionless heat flux}
\textbf{Subscripts}

\entry{$1$}{bottom wall ``1''}
\entry{$2$}{top wall ``2''}
\entry{$0$}{point at which the shear stress is zero}
\entry{$w$}{wall}
\entry{$m$}{mean}
\end{nomenclature}


\bibliographystyle{asmems4}
\bibliography{asymmetric-slip-plaw.bib}


\appendix

\section{Explicit expressions for some integrals}
\label{app:integrals}

Although it can be accomplished by hand, using {\sc Mathematica}, we find from Eq.~\eqref{velocity_nd} that the dimensionless flow rate $\bar{q}$ in the channel is
\begin{multline}
\bar{q} = \int_0^1 \bar{u}(\bar{y}) \, d\bar{y} = 
\frac{1}{2n+1}\Bigg\{ (1-\bar{\tau}_{w1})^{1/n} \\ 
- (2 n+1) (\bar{\tau}_{w1}-1) \bar{u}_{w2}\\
+ \bar{\tau}_{w1} \left[\bar{\tau}_{w1}^{{1}/{n}+1} + (\bar{\tau}_{w1} -2) (1 - \bar{\tau}_{w1})^{1/n} + \bar{u}_{w1} \right]\\ 
+ n \left[\bar{\tau}_{w1} ^{{1}/{n}+2}+(1-\bar{\tau}_{w1} )^{{1}/{n}+2}+2 \bar{\tau}_{w1} \bar{u}_{w1}\right] \Bigg\}.
\end{multline}
As before, $\bar{u}_{w1}$ and $\bar{u}_{w2}$ are given by Eqs.~\eqref{velocity_nd_bc}, while $\bar{\tau}_{w1}$ is the solution to Eq.~\eqref{velocity_nd_tau_eq}. 
Similarly, it can be shown that
\begin{multline}
\int_0^1\left|\frac{d\bar{u}}{d\bar{y}}\right|^{n-1} \left(\frac{d\bar{u}}{d\bar{y}}\right)^2 d\bar{y}\\
= \frac{\left({1}/{n}+1\right)^n (n+1)}{2 n+1} \left[ \bar{\tau}_{w1}^{{1}/{n}+2}+(1-\bar{\tau}_{w1})^{{1}/{n}+2} \right].
\end{multline}

\end{document}